\documentclass[prl,amsmath,amssymb,reprint,onecolumn]{revtex4-1}
\usepackage{CJK}
\usepackage{graphicx}% Include figure files
\usepackage{dcolumn}% Align table columns on decimal point
\usepackage{bm}% bold math
\usepackage[mathlines]{lineno}% Enable numbering of text and display math
\usepackage[utf8]{inputenc}
\usepackage[T1]{fontenc}
\usepackage{mathptmx} 

\usepackage[dvipsnames]{xcolor} %leo
\usepackage{soul}
\begin{document}

\title{Pore-shape and its spatial organization control intrinsic permeability of porous media}

\author{Wenqiao Jiao}
\affiliation{Institute of Earth Science, University of Lausanne, Lausanne 1015, Switzerland}
\affiliation{Department of Civil and Environmental Engineering, Massachusetts Institute of Technology, Cambridge, MA 02139, USA}

\author{Isaac Pincus}
\affiliation{Institute of Earth Science, University of Lausanne, Lausanne 1015, Switzerland}

\author{Chiara Recalcati}
\affiliation{Department of Water Resources and Drinking Water, Eawag, Swiss Federal Institute of Aquatic Science and Technology, Dübendorf 8600, Switzerland}

\author{Alberto Guadagnini}
\affiliation{Department of Civil and Environmental Engineering, Politecnico di Milano, Milano 20133, Italy}
\affiliation{Sonny Astani Department of Civil and Environmental Engineering, Viterbi School of Engineering, University of Southern California, Los Angeles, CA 90089-2531, USA}

\author{Pietro de Anna}
\email{pietro.deanna@unil.ch}
\affiliation{Institute of Earth Science, University of Lausanne, Lausanne 1015, Switzerland}

\date{\today}

\begin{abstract}
\noindent
The structure of a porous material, and in particular its spatial variability, is known to control the intrinsic permeability of the system. We investigate how dead-end pores influence the intrinsic permeability of a porous medium beyond their contribution to total pore volume. Dead-end pores are ubiquitous in porous media, yet they are often treated as hydraulically inactive regions whose influence is assumed to be negligible or absorbed into effective-porosity descriptions. We perform pore-scale flow simulations across different dead-end pore structures, including heterogeneous arrangements, controlled granular assemblies, and a minimal single-channel model to study their impact on the system macroscopic permeability. This strategy allows us to isolate the effects of dead-end pore density, depth, and orientation while preserving the transmitting network. We find that dead-end pores can influence intrinsic permeability: increasing the density of dead-end pores along percolating flow paths enhances permeability, whereas pore depth and junction orientation have negligible effects. The observed permeability enhancement originates from localized hydrodynamic interactions at junctions between transmitting and dead-end pores. Based on these results, we propose an effective formulation that relates the density and spatial organization of dead-end pores relative to the transmitting network to macroscopic permeability. Our findings show that dead-end pore architecture provides an additional geometric control on intrinsic permeability beyond porosity and pore-size statistics.
\end{abstract}

\maketitle
%\linenumbers

\section{Introduction}

\noindent
In porous systems characterized by a complex and intricate internal architecture, fluid flow is typically self-organized through preferential pathways that tend to bypass regions of low velocity or stagnation. Within such systems, fluid flow is governed by the heterogeneous geometry of the pore space at the microscale ~\citep{deAnnaPRF2017, alim2017local}. Detailed understanding of this pore-scale control is therefore key for accurate quantification of fluid flow through porous materials and underpins a wide range of natural and engineered scenarios, including deep geothermal energy production~\cite{watanabe2020stabilizing}, geological ${CO_2}$ storage~\cite{de2016influence}, remediation of contaminated soil-water systems~\cite{qiu2021experiments}, and riverbank filtration~\cite{banzhaf2011investigative}. At the macroscopic level, fluid motion in porous media under laminar conditions is commonly described by Darcy’s law, $q = -\frac{k}{\mu}\nabla P,$ where $q$ is the average fluid velocity, $\mu$ is the dynamic viscosity of fluid, $\nabla P$ is the macroscopic pressure gradient, and $k$ denotes the intrinsic permeability of the medium~\cite{bear1988dynamics}. The latter is a key effective hydraulic property that characterizes the ability of a porous material to transmit fluids under an applied pressure gradient. As such, it reflects the integrated effect of the pore-space structure and its organization across multiple length scales~\cite{graczyk2020predicting, luo2014numerical, liu2019pore, sperry1995model, cormican2020grain, kang2014pore,jiao2024intrinsic}.\\

\noindent
While classical models such as the Kozeny-Carman formulation~\cite{kozeny1927kapillare, carman1937fluid}, $k = \frac{d_g^2}{180} \frac{\phi^3}{(1 - \phi)^2}$, can provide reliable permeability estimates for relatively uniform porous materials, their reliance on bulk-averaged quantities (e.g., grain size $d_g$, porosity $\phi$, and tortuosity $\tau$) limits their ability to adequately represent the intricate pore-scale organization of heterogeneous media~\cite{hommel2018porosity, schulz2019beyond}. This limitation is especially pronounced in natural porous systems, which commonly exhibit broad, multiscale pore-size distributions spanning several orders of magnitude~\cite{bear1988dynamics, blunt2013pore, da2021deep}. In such materials, pore diameters may range from micrometers (10$^{-6}$~m) to centimeters (10$^{-1}$~m) and are frequently characterized by power-law or log-normal distributions~\cite{gueven2017hydraulic, mizutani2010relation, zhao2017integrating}.
In this context, substantial research efforts have been focused on improving the quality of permeability estimates upon extending the Kozeny-Carman formulation to media composed of non-uniform grains or exhibiting other forms of structural heterogeneity. These approaches are designed to incorporate descriptors such as pore-size distributions, effective grain size, characteristic length scales, and porosity~\cite{hommel2018porosity, koponen1997permeability, yang2014analytical, nishiyama2017permeability, schulz2019beyond, sheikh2015numerical}. Nevertheless, many of these formulations remain limited by their reliance on a single (average or representative) pore throat and/or grain size. Moreover, some of the empirical (or fitted) parameters embedded in these models have minimal (or no) direct physical correspondence to the actual pore structure of the medium and cannot be quantified through direct measurement or observation~\cite{xu2008developing, hyman2013pedotransfer}. \\

\noindent
We recently introduced a theoretical modeling framework~\cite{jiao2024intrinsic} that explicitly incorporates pore-size variability in heterogeneous porous media. Within this approach, the intrinsic permeability $k$ is derived analytically based solely on structural information of the porous medium. The porous structure is conceptualized as a series of $m = L \tau / \overline{w}$ elementary porous units arranged along the mean flow direction (here, $L$ denotes the macroscopic length of the medium and $\overline{w}$ is the average pore body size). Each $i$-th unitary pore is idealized as a cylindrical conduit, whose permeability $k_i$ can be obtained via Hagen-Poiseuille law~\cite{sutera1993history}. Under this formulation, the macroscopic permeability $k$ is expressed in terms of the distribution of pore-body sizes $w_i$ and pore-throat sizes $\lambda_i$ as $k =L\,\tau^2  /  (\sum_{i=1}^{m} \frac{32\, w_i}{\lambda_i^2})$. Notably, this formulation yields accurate estimates of intrinsic permeability without introducing any empirical fitting parameters. Consequently, when the spatial arrangement of pores is explicitly resolved, pore-size variability naturally emerges as a primary control on intrinsic permeability\\

\noindent
However, pore-size heterogeneity constitutes only one component of the broader pore-scale heterogeneity characteristic of natural porous media, which also arises from the morphological complexity associated with disordered pore structures~\cite{alhashmi2016impact,bear2012hydraulics,dupont2011acoustic,wu2019predicting}. In addition to differences in pore size, realistic porous media exhibit pronounced geometric and topological complexity, including variations in pore shape, connectivity, and local structural organization~\cite{lever1985effect,bear2012hydraulics,dupont2011acoustic,fatt1961influence}. From a hydrodynamic perspective, advective transport is sustained primarily by the connected portion of the pore space that actively conveys flow, commonly referred to as transmitting pores (TPs)~\cite{bordoloi2022structure}. In contrast, non-transmitting pores (hereafter termed dead-end pores (DEPs)) do not directly contribute to the net flow. Yet it remains unclear whether dead-end pores influence intrinsic permeability solely through their contribution to total pore volume, or whether their spatial organization relative to the transmitting network provides an additional independent structural control. Notably, dead-end pores are typically linked to the transmitting backbone through narrow single throats~\cite{bear2012hydraulics,dupont2011acoustic}. Although such features are widespread in many permeable systems such as rocks~\cite{hemingway1983effect, soltanmohammadi2024investigation,zou2018impacts}, soils~\cite{nishiyama2017permeability}, biological tissues~\cite{hapfelmeier2010reversible,jiao2025spatial,scheidweiler2024spatial}, and polymeric materials~\cite{phillip2011tuning}, the role of dead-end pore morphology in permeability modeling frameworks remains insufficiently understood. \\

\noindent
Here, we explore how DEPs affect intrinsic permeability beyond their mere contribution to total pore volume. To this end, we rely on a simulation-based strategy and analyze several classes of pore assemblies with systematically controlled levels of structural variability. We first consider highly heterogeneous pore structures representative of complex natural media, where TPs coexist with randomly distributed and morphologically diverse DEPs. We then design idealized controlled granular media in which the density of DEPs relative to the transmitting network is systematically varied while preserving the structure of the conducting network. Finally, we introduce a minimal single-pore model that captures the essential geometric features of the system while allowing independent adjustment of DEP depth, DEP density, and the orientation of TP-DEP junctions. By independently tuning these geometric characteristics and evaluating the corresponding permeabilities, we isolate the structural features of DEPs that govern macroscopic flow resistance. Our findings reveal that intrinsic permeability cannot be determined solely from bulk porosity or pore-size distributions. Rather, it also depends strongly on the density and spatial arrangement of DEPs with respect to the transmitting network.\\

\section{Numerical framework and validation}

\begin{figure}[h!]
    \begin{center}
        \includegraphics[trim={0 0 0 0},clip,width=1\textwidth]{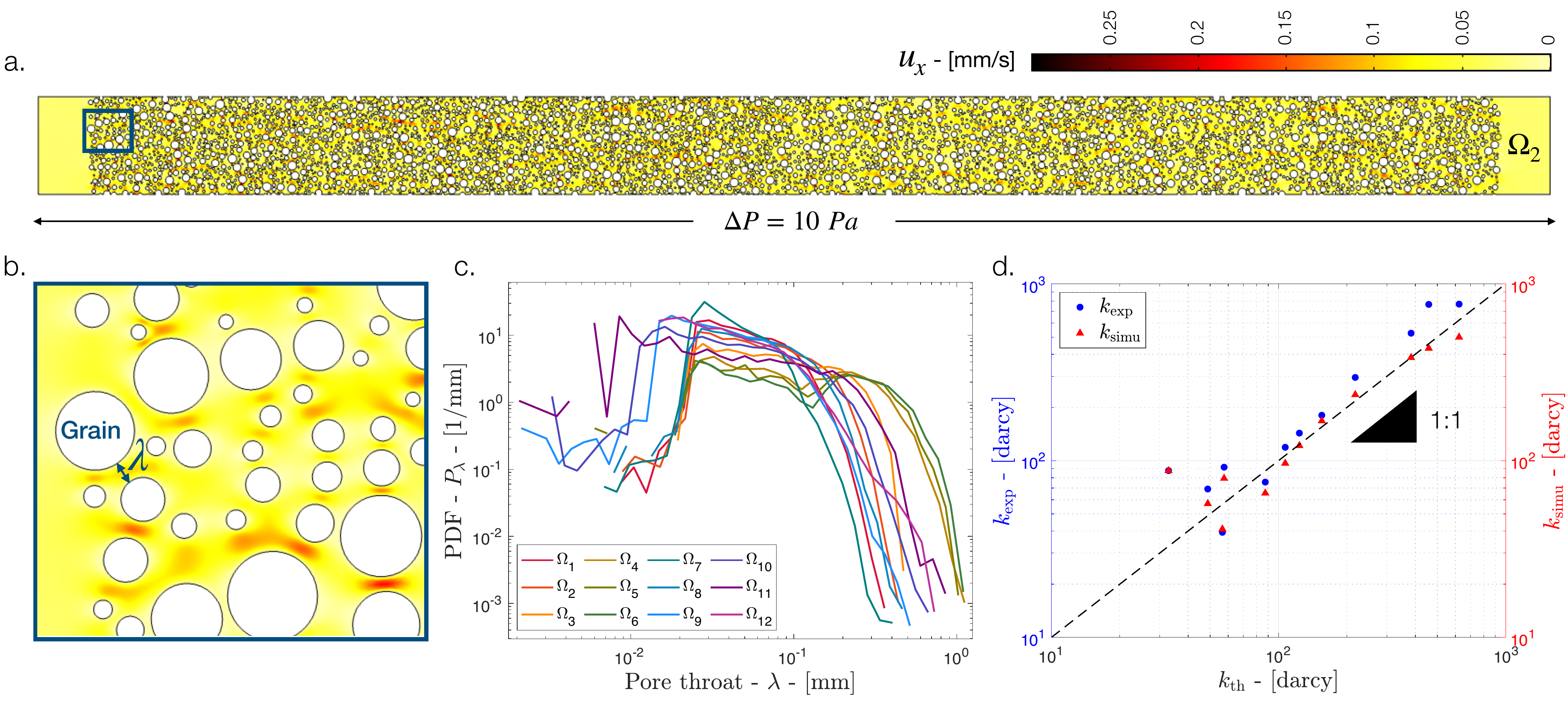} \\
       \caption{Validation of the numerical framework against experimental and theoretical results.
            (a) Representative quasi-two-dimensional porous geometry $\Omega_2$ selected from the set of 12 microfluidic samples reported in Ref.~\cite{jiao2024intrinsic}. 
            (b) Enlarged view of the inlet region highlighted in (a); white circles denote solid grains, the distance between two neighboring grains defining the local pore throat $\lambda$.
            (c) Log-log plot of the probability density function (PDF), $ P_{\lambda}$, of pore-throat sizes for the 12 heterogeneous geometries used in the validation.
            (d) Comparison of intrinsic permeability obtained from numerical simulations ~($k_{\mathrm{simu}}$), microfluidic experiments ($k_{\mathrm{exp}}$), and the theoretical model introduced in Ref.~\cite{jiao2024intrinsic} ($k_{\mathrm{th}}$). The dashed line corresponds to the 1:1 relation.}  \label{Figure1}
    \end{center}
\end{figure}

\noindent
We perform numerical simulations of steady, incompressible laminar flow through quasi-two-dimensional (hereafter termed 2.5D) porous geometries using \textit{COMSOL Multiphysics}. We start by validating the numerical simulation scheme on the basis of the experimental and theoretical results reported in our previous work~\cite{jiao2024intrinsic}. To this end, we consider the 12 reference microfluidic porous structures introduced in Ref.~\cite{jiao2024intrinsic} (corresponding to configurations herein denoted as $\Omega_1 - \Omega_{12}$). Figures~\ref{Figure1}~(a,b) depict a selected configuration (termed $\Omega_2$), together with the associated depth-averaged velocity field. This geometry is constructed from a collection of non-overlapping disks with random sizes and spatial distributions. It does not include dead-end pores. The full set of 12 geometries spans a broad range of pore-throat sizes $\lambda$, whose sample probability density functions (PDFs) are depicted in Fig.~\ref{Figure1}~(c).\\

\noindent
At the macroscopic scale, flow is driven by a constant pressure drop ($\Delta p$) imposed between the inlet and outlet. No-slip boundary conditions are imposed on all solid surfaces. To account for the finite thickness of the microfluidic channel ($h = 0.1$~mm~\cite{jiao2024intrinsic}), we adopt a 2.5D depth-averaged formulation~\cite{deng20162, roman2016particle, soulaine2021computational}, according to which the in-plane velocity field $\mathbf{u}$ and pressure $p$ satisfy the steady-state Brinkman equations for incompressible fluid
\begin{equation}
    \mathbf{0} = \nabla \cdot \left(-p \mathbf{I} + \mu \left(\nabla \mathbf{u} + (\nabla \mathbf{u})^{T}\right)\right) + \mathbf{F}, \qquad \nabla \cdot \mathbf{u} = 0.
\end{equation}
The finite channel thickness introduces an additional viscous dissipation in the out-of-plane direction. The latter is accounted for through an effective linear drag term of Hele--Shaw type~\cite{lamb1924hydrodynamics}. This contribution can be interpreted as a Darcy-like resistance~\cite{roman2016particle} associated with an effective permeability $k_h=h^2/12$, thus yielding

\begin{equation}
    \mathbf{F} = -\frac{12 \mu}{h^{2}} \mathbf{u}.
\end{equation}

\noindent
Intrinsic permeability is evaluated from the simulated volumetric flux $Q$ using Darcy's law (i.e., $k=-\mu\frac{Q/A}{\Delta p/L}$, where $A$ is the cross-sectional area normal to the mean flow direction and $L$ is the length of the porous medium). In all simulations, the computational mesh is refined to ensure that the smallest pore throat is resolved by at least 10 elements, thus ensuring sufficient spatial resolution for permeability evaluation. As shown in Fig.~\ref{Figure1}~(d), numerically obtained permeabilities are in excellent agreement with both experimentally measured values and analytical predictions included in  Ref.~\cite{jiao2024intrinsic} across the full set of heterogeneous geometries. This consistency is a strong element imbuing us with confidence of the appropriateness of the numerical framework which is here employed to investigate the impact of DEPs on intrinsic permeability. \\

\section{Heterogeneous porous media with dead-end pores}

\noindent
To generate a realistic porous structure characterized by the presence of DEPs, we employ a Cahn-Hilliard phase-field approach~\cite{bartels2021cahn}. This method yields grain-like solid regions embedding cavities (i.e.,  DEPs) that are separated by a connected network of TPs. The characteristic size of the TP network is quantified through the Maximum Inscribed Circle (MIC) method~\cite{bordoloi2022structure, jiao2024intrinsic}, yielding an average TP size $\overline{D}_{\mathrm{MIC}} = 81~\mu$m. The porous domain is characterized by fixed macroscopic dimensions corresponding to $L = 49.75$~mm and $W = 3.98$~mm. All numerical simulations are performed using the depth-averaged framework described in the previous section, with a channel height $h = 100~\mu$m.\\
\begin{figure}[h!]
    \begin{center}
    \includegraphics[trim={0 0 0 0},clip,width= 1\textwidth]{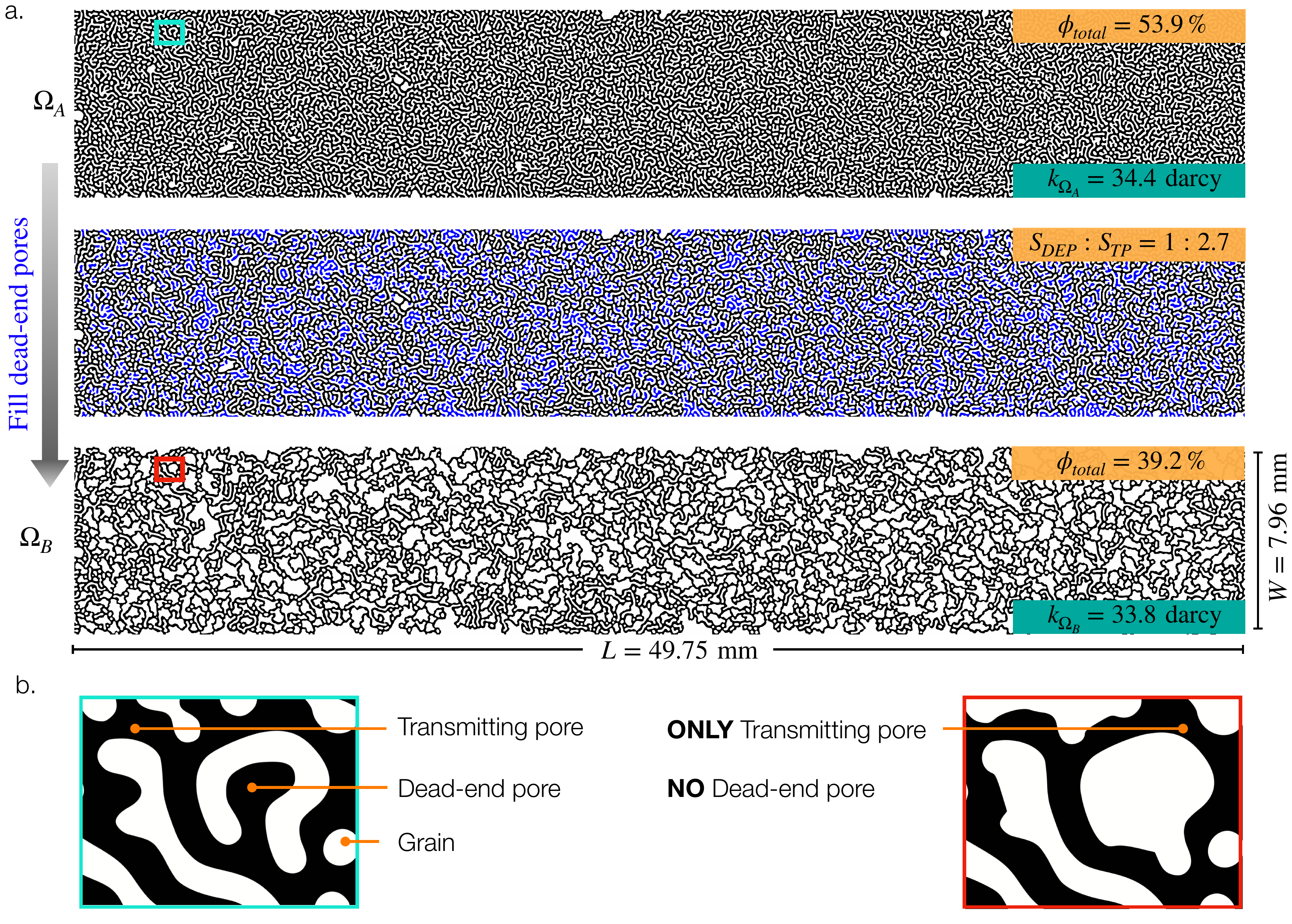} \\
    \caption{Pore geometries with and without dead-end pores. (a) Phase-field--generated pore structures with identical macroscopic dimensions and distinct pore connectivity. Configuration $\Omega_A$ (top) features a connected network of transmitting pores (TPs) coexisting with numerous dead-end pores (DEPs). The reference configuration $\Omega_B$ (bottom) is obtained by filling all DEPs with solid material (blue), yielding a pore space composed solely of transmitting pores. (b) Local magnification of the pore space in $\Omega_A$ and $\Omega_B$, highlighting geometric differences between dead-end and transmitting pores.}\label{Figure2}
    \end{center}
 \end{figure}

\noindent
We consider the system shown in Fig.\ref{Figure2}~(a). The structure $\Omega_A$ is obtained from a phase-field-generated structure characterized by a percolating network of TPs coexisting with DEPs (see Fig.~\ref{Figure2}~(a), top). The total porosity of $\Omega_A$ is $\phi_{total} = 53.9\%$, with a DEP-to-TP volume ratio of $S_{DEP} : S_{TP} = 1 : 2.7$. Numerical simulations yield an intrinsic permeability of $k = 34.4$~darcy.\\

\noindent
To assess the contribution of DEP volume to intrinsic permeability, we construct a modified configuration $\Omega_B$ (Fig.~\ref{Figure2} (a), bottom) by filling all DEPs in $\Omega_A$ with solid material, thereby effectively converting them into grains (Fig.~\ref{Figure2} (b)). This procedure modifies the original macroscopic geometry ($\Omega_A$) and leaves the transmitting network unchanged, while reducing the total porosity to $\phi_{total} = 39.2\%$. The ensuing intrinsic permeability is $k = 33.8$~darcy. The imposed reduction in porosity, exceeding 14 percentage points, leads to a permeability decrease of less than 2\%. This relatively small decrease in permeability is associated with different boundary conditions for fluid flow at the DEP-TP junction. Within TPs, the fluid experiences viscous dissipation similar to a pipe system, i.e., drag forces at the solid walls. Otherwise, at a TP-DEP junction, the fluid flowing in TPs somehow interacts with the laminar vortex driving local flow field within DEP~\cite{bordoloi2022structure, chuanfeng2021effects}. The observed sensitivity is in contrast with conventional porosity-based permeability models and suggests that DEP volume alone does not govern intrinsic permeability~\cite{koponen1997permeability}. \\
\begin{figure}[h!]
\begin{center}
    \includegraphics[trim={0 0 0 0},clip,width=1\textwidth]{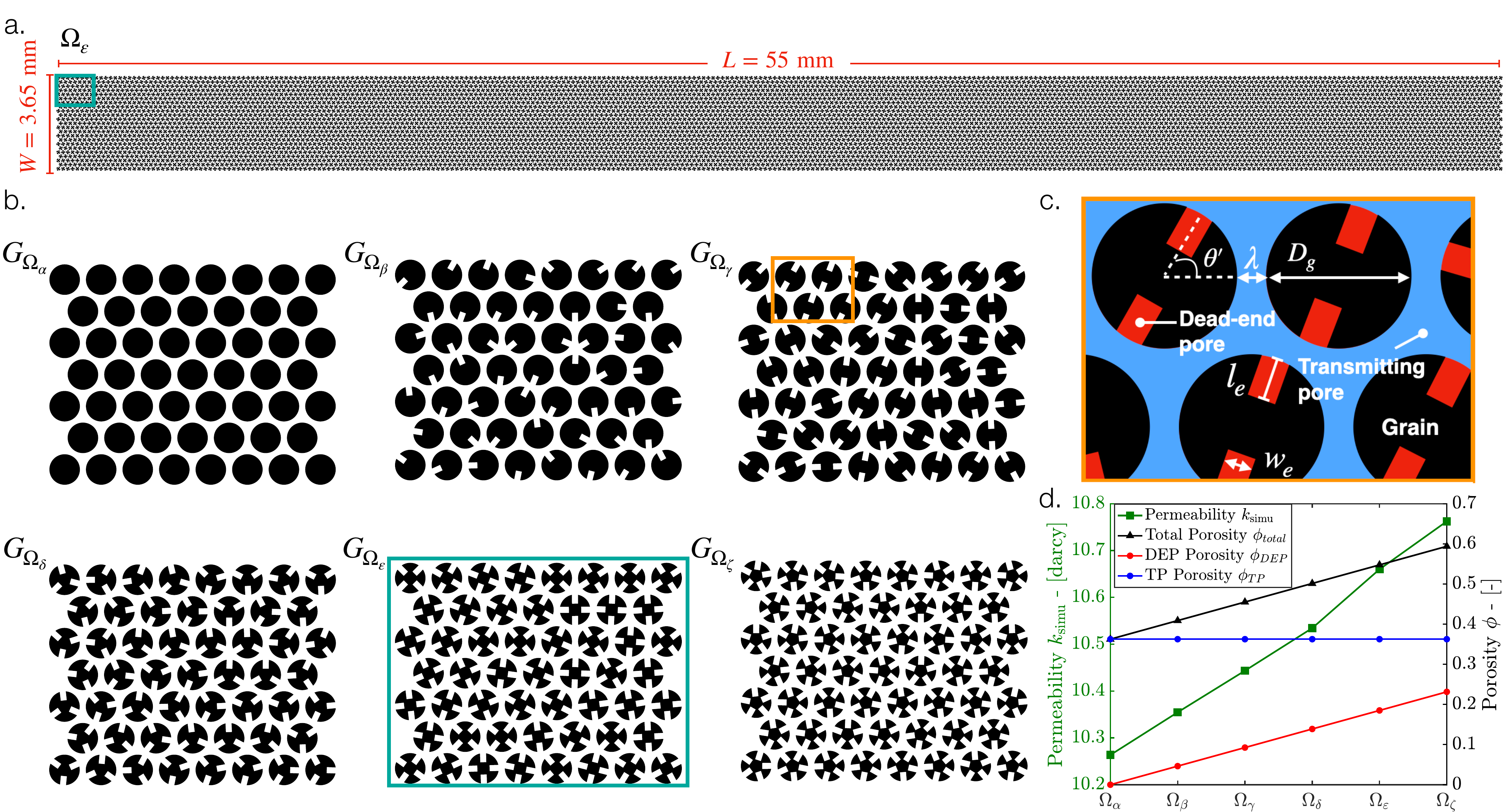}
    \caption{Porous media with controlled dead-end pore density. (a) Representative configuration $\Omega_5$, consisting of a homogeneous packing of monodisperse grains containing multiple dead-end pores per grain (domain dimensions are $L = 55$~mm and $W = 3.65$~mm). (b) Schematic pore patterns $G_{\Omega_{\alpha}}$--$G_{\Omega_{\zeta}}$ corresponding to configurations with $n_e = 0$--$5$ dead-end pores per grain. (c) Local magnification of $G_{\Omega_{\gamma}}$ illustrating the geometric parameters of the system, including grain diameter ($D_g$), dead-end pore width ($w_e$), depth ($l_e$), and orientation of the TP-DEP junction relative to the mean flow direction. (d) Total porosity ($\phi_{total}$), transmitting porosity ($\phi_{TP}$), and intrinsic permeability ($k$) obtained from numerical simulations for  $\Omega_{\alpha}$--$\Omega_{\zeta}$. While $\phi_{TP}$ remains constant, increasing the number of dead-end pores leads to a permeability enhancement of approximately 5\%.}  \label{Figure3}
\end{center}
\end{figure}

\noindent
To further investigate the way medium permeability depends on the spatial organization and geometric properties of DEPs relative to the transmitting network, we design a family of porous systems in which selected DEP characteristics are systematically varied while keeping the architecture of TPs. Specifically, we consider configurations with ($i$) varying ratios of dead-end to transmitting pores, ($ii$) controlled DEP depths, and ($iii$) systematically varied spacing and density of DEPs along representative flow paths. \\

\section{Porous Media with Controlled Dead-End Pore Density}

\noindent
To isolate the influence of DEPs from bulk porosity effects, we construct a series of idealized porous media in which the transmitting pore network is preserved, while the number of DEPs encountered along each inlet-to-outlet flow path is systematically varied. The resulting configurations, denoted as $\Omega_{\alpha}$--$\Omega_{\zeta}$, are depicted in Fig.~\ref{Figure3}.\\

\noindent
All six structures are based on a homogeneous packing of monodisperse, non-overlapping circular grains with diameter $D_g = 150~\mu$m, arranged within a domain of fixed dimensions ($L = 55$~mm and $W = 3.65$~mm). Configuration $\Omega_{\alpha}$ is designed to contain only TPs. Thus, it is taken as reference. In configurations $\Omega_{\beta}$--$\Omega_{\zeta}$, each grain is modified to include $n_c = 1-5$ cavities acting as DEPs (see Fig.~\ref{Figure3}~(b)). Each cavity is modeled as a rectangular opening with width $w_e = 30~\mu$m and depth $l_e = 45~\mu$m (Fig.~\ref{Figure3}~(c)). Grain orientations are randomized throughout the domain, resulting in a uniform distribution of DEP junction orientations relative to the mean flow direction.  \\

\noindent
This construction increases the DEP volume fraction ($\phi_{DEP} = V_{DEP} / V_{total}$) while maintaining the same TP volume fraction ($\phi_{TP} = V_{TP} / V_{total}$) across all configurations. As a result, the total porosity ($\phi_{total} = (V_{DEP} + V_{TP})/ V_{total})$ increases monotonically with the number of cavities per grain (i.e., $n_c$), as shown in Fig.~\ref{Figure3}~(d). These geometries therefore provide a controlled framework to assess the impact of DEP density independently of the transmitting structure. \\

\noindent
Numerical simulations of fluid flow through these structures reveal that permeability systematically varies with DEP density. As shown in Fig.~\ref{Figure3}~(d), increasing the number of cavities per grain (and thus the density of DEPs along flow paths) leads to an increase in permeability of up to 5~\% compared to the DEP-free scenario (i.e., $\Omega_{\alpha}$). To elucidate the origin of this observed behavior, we next analyze the hydraulic response at the scale of an individual DEP-TP junction.\\

\section{Single-pore model to isolate the DEP effect on permeability}

\begin{figure}[h!]
\begin{center}
    \includegraphics[trim={0 0 0 0},clip,width=1\textwidth]{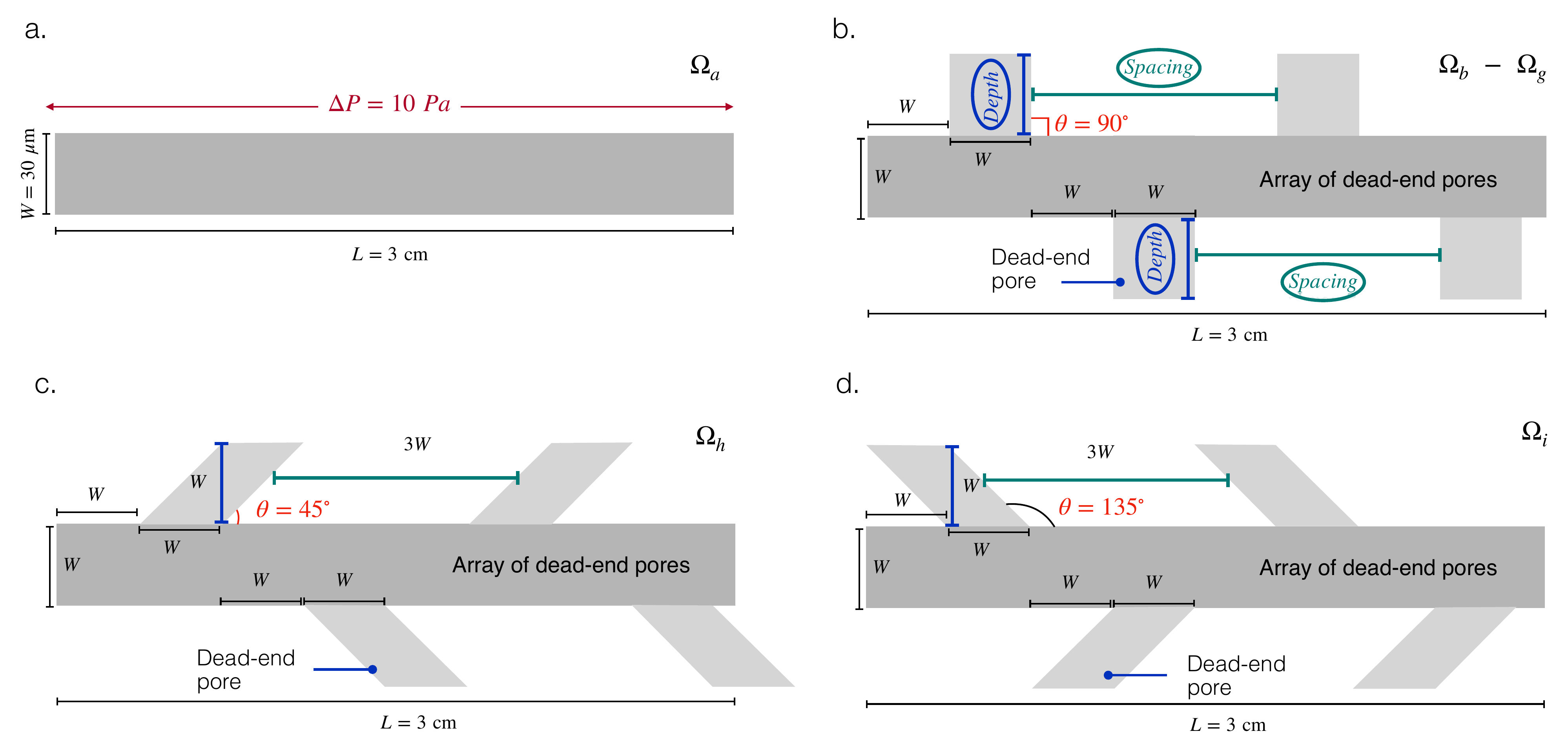} \\
    \caption{Single-channel configurations with controlled dead-end pore geometry.
(a) Reference straight channel $\Omega_a$ without dead-end pores. 
(b) Parameterized geometry for configurations $\Omega_b$--$\Omega_g$, illustrating dead-end pores of prescribed depth and spacing, with junction orientation $\theta$ ($\theta=90^\circ$). 
(c) Inclined dead-end pore configuration $\Omega_h$ with $\theta=45^\circ$. 
(d) Inclined dead-end pore configuration $\Omega_i$ with $\theta=135^\circ$.}  \label{Figure4}
\end{center}
\end{figure}

\noindent
To isolate the hydrodynamic influence of dead-end pores on the porous system permeability, we model a TP as a straight channel of width $W$, to which cavities (DEPs) of prescribed depth, spacing (i.e. linear density), and orientation are systematically added. Overall, we consider nine configurations, denoted as $\Omega_a$--$\Omega_i$ ($\Omega_a$ corresponding to a reference DEP-free scenario). In all cases, the main channel has length $L = 3$ cm and width $W = 30~\mu$m, and it is subject to a constant pressure drop $\Delta P = 10$~Pa. These are schematically depicted in Fig.~\ref{Figure4} and their properties are summarized in Table~\ref{tab:single_pore_configs}.\\

\begin{table}[htbp]
\centering
\small
\caption{Geometric parameters and intrinsic permeability of single-channel configurations. The dead-end pore depth, spacing, number, junction orientation, and the corresponding simulated intrinsic permeability are listed for each configuration.}
\label{tab:single_pore_configs}
\renewcommand{\arraystretch}{1.4}
\setlength{\tabcolsep}{13pt}
\begin{tabular}{lccccccccc}
\hline
\textbf{Geometry}
& $\boldsymbol{\Omega_a}$ & $\boldsymbol{\Omega_b}$ & $\boldsymbol{\Omega_c}$ & $\boldsymbol{\Omega_d}$
& $\boldsymbol{\Omega_e}$ & $\boldsymbol{\Omega_f}$ & $\boldsymbol{\Omega_g}$ & $\boldsymbol{\Omega_h}$ & $\boldsymbol{\Omega_i}$ \\
\hline
Depth
& $0$ & $W$ & $2W$ & $4W$ & $W$ & $W$ & $W$ & $W$ & $W$ \\

Spacing
& $0$ & $3W$ & $3W$ & $3W$ & $6W$ & $9W$ & $21W$ & $3W$ & $3W$ \\

Number of dead-end pores
& $0$ & $500$ & $500$ & $500$ & $286$ & $200$ & $91$ & $500$ & $500$ \\

Orientation $\theta$
& $0$ & $90^\circ$ & $90^\circ$ & $90^\circ$ & $90^\circ$ & $90^\circ$ & $90^\circ$ & $45^\circ$ & $135^\circ$ \\

$k_{\mathrm{simu}}$ [darcy]
& $34.1$ & $37.3$ & $37.3$ & $37.3$ & $35.9$ & $35.3$ & $34.7$ & $37.2$ & $37.1$ \\
\hline
\end{tabular}
\end{table}

\noindent
A direct comparison between $\Omega_a$ and $\Omega_b$, where DEPs of depth $W$ are introduced at a fixed spacing of $3W$, reveals that the presence of DEPs increases intrinsic permeability by approximately 9\% relative to the scenario without DEPs ($\Omega_a$). To isolate the effect of DEP depth, configurations $\Omega_b$, $\Omega_c$, and $\Omega_d$ are constructed with identical DEP spacing, while varying their depths to $W$, $2W$, and $4W$, respectively. Despite this variation, the intrinsic permeability remains unchanged, indicating a negligible sensitivity to DEP depth. \\

\noindent
In contrast, the density of DEPs along flow paths exerts a pronounced control on permeability. Configurations $\Omega_b$, $\Omega_e$, $\Omega_f$, and $\Omega_g$ differ only in the spacing between adjacent DEPs (set to $3W$, $6W$, $9W$, and $21W$, respectively), while keeping DEP depth as constant. As spacing increases (i.e., DEP linear density decreases), the intrinsic permeability progressively reduces and attains a value similar to $k_{\Omega_a}$ (i.e., purely TP scenario). This behavior can be attributed to the interplay between no-slip conditions along solid wall boundaries and fluid-fluid interfaces at TP-DEP junctions, which, taken together, modulate the effective hydraulic resistance. \\

\noindent
Finally, the influence of DEP-TP junction orientation is taken into consideration by comparing configurations $\Omega_h$, $\Omega_b$, and $\Omega_i$, in which the DEPs are tilted  relative to the mean flow direction at $\theta=45^\circ$, $90^\circ$, and $135^\circ$, respectively, while fixing DEP depth and spacing. The resulting permeability values remain nearly unchanged, indicating that, within the present linear Brinkman framework, permeability enhancement is largely unaffected by junction orientation. In particular, the close agreement between $\Omega_h$ and $\Omega_i$ is consistent with the reversibility of low-Reynolds-number flows for mirror-symmetric configurations.\\

\begin{figure}[h!]
\begin{center}
    \includegraphics[trim={0 0 0 0},clip,width=1\textwidth]{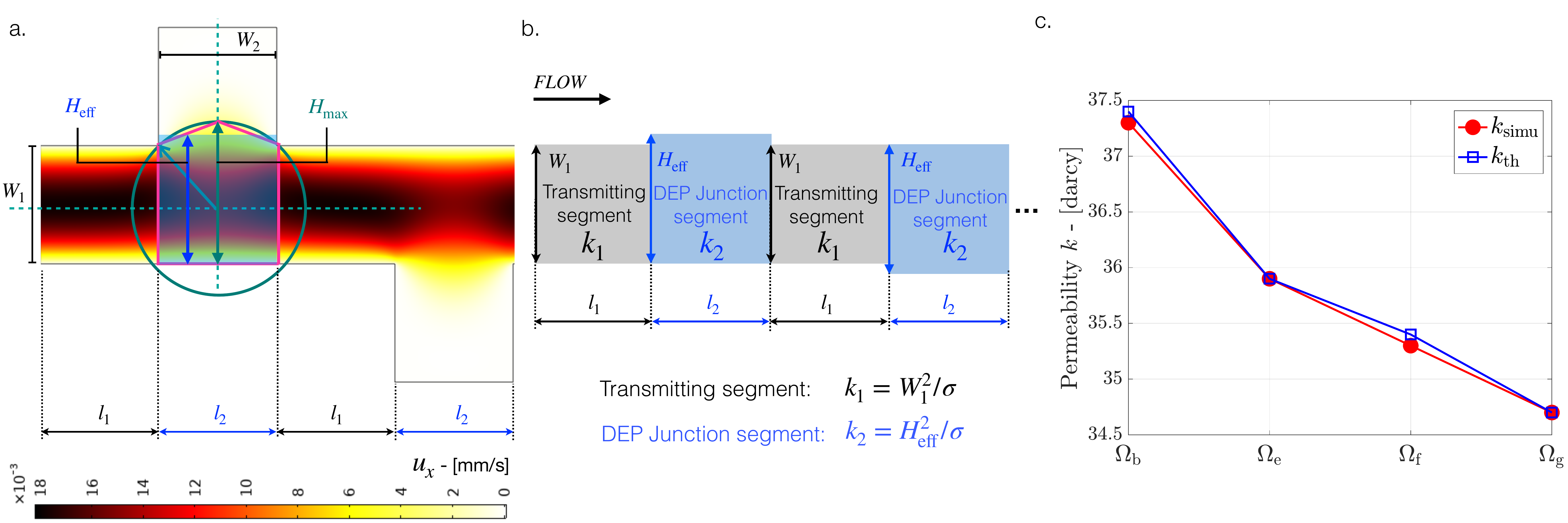}
    \caption{Junction-scale control on permeability and effective resistance model. (a) Magnification of the pore-scale velocity field $u_x$ in a representative portion of configuration $\Omega_b$, highlighting the confined hydrodynamic interaction at a TP-DEP junction. The junction is characterized by an effective aperture $H_{\mathrm{eff}}$, defined from the local junction geometry through Eqs.~(\ref{eq:Heff}) and (\ref{eq:Hmax}). (b) Effective streamwise (one-dimensional) representation of the channel as a sequence of segments of equal length $W$ (i.e., $l_1=l_2=W$), with distinct hydraulic responses: transmitting segments with $k_1=W_1^2/\sigma$ and DEP-junction segments with $k_2=H_{\mathrm{eff}}^2/\sigma$. For increasing DEP spacing, multiple consecutive transmitting segments represent the region between adjacent junctions. (c) Comparison between theoretically-based permeability values $k_{\mathrm{th}}$ and their numerical counterparts $k_{\mathrm{simu}}$ for configurations $\Omega_b$, $\Omega_e$, $\Omega_f$, and $\Omega_g$. The excellent agreement demonstrates that permeability variations are governed by junction-scale hydrodynamic effects captured by the effective-aperture model.}
\label{Figure5}
\end{center}
\end{figure}

\noindent
Taken together, these findings suggest that permeability variations induced by the presence of DEPs originate from localized hydrodynamic interactions at  TP-DEP junctions, rather than from DEP depth or junction orientation. Fig.~\ref{Figure5}~(a) illustrates the velocity field at a TP-DEP junction for configuration $\Omega_b$ and reveals the presence of confined interaction regions near the DEP junction. These, in turn, alter the hydraulic conductance of the transmitting channel. Motivated by this observation, we conceptualize the channel as composed by TP and DEP-TP junction segments (see Fig.~\ref{Figure5}~(b)).\\

\noindent
Effectively, each DEP-TP junction locally increases the pore size and, thus, the individual pore permeability. Thus, by properly quantifying the pore size distribution (e.g. with a MIC algorithm) it is possible to assess the macroscopic system permeability with the model described in Ref.~\cite{jiao2024intrinsic}. This model conceptualizes the whole porous landscape as a sequence of $m$ smaller porous systems in series of length $l_i$ and permeability $k_i$: the system permeability $k_{th}$ will be, then, the harmonic mean of all media permeability $k_i$, weighted by their own extent $l_i$. We analytically estimate the system permeability by decomposing it into streamwise segments: a number $n_1$ of them will be a TP of extent $W_1$, while $n_2$ will be a DEP-TP junction of extent $W_2$. Thus, $n_1$ and $n_2$ must satisfy
\begin{equation}
    W_1 n_1+ W_2 n_2 = L.
\end{equation}
and the total system permeability would be
\begin{equation}
    \frac{1}{k_{\mathrm{th}}} = \frac{1}{L} \left( n_1\,\frac{W_1}{k_1} + n_2\,\frac{W_2}{k_2} \right).
\label{eq:kth_resistance}
\end{equation}

\noindent
The intrinsic permeability associated with each segment type is defined as
\begin{equation}
    k_1=\frac{W_1^{2}}{\sigma}, \quad k_2 = \frac{H_{\mathrm{eff}}^{2}}{\sigma},
\end{equation}
where $H_{\mathrm{eff}}$ is the effective DEP-junction aperture. For the particular case analyzed here, we define the effective DEP-TP junction aperture as the average between the TP aperture $W_1$ and the radius of the circle represented in Fig.~\ref{Figure5}~(a):
\begin{equation}
    H_{\mathrm{eff}}=\frac{W_1+H_{\max}}{2},
\label{eq:Heff}
\end{equation}
with
\begin{equation}
    H_{\max} = \sqrt{\left(\frac{W_1}{2}\right)^{2}+\left(\frac{W_2}{2}\right)^{2}}+\frac{W_1}{2}.
\label{eq:Hmax}
\end{equation}
In general, $H_\mathrm{eff}$ can be determined with a proper analysis of the porous structure, for instance with a MIC algorithm~\cite{jiao2024intrinsic}. We expressed the permeability of each segment with the geometry-dependent parameter $\sigma$ linking the characteristic aperture to hydraulic conductance. For canonical geometries, this parameter is known analytically. For example, for a circular pipe of diameter $D$, the Hagen--Poiseuille solution ~\cite{sutera1993history} yields $k=D^2/32$, so that $\sigma = 32$. For rectangular channels of height $H^{\prime}$ and width $W^{\prime}$, the solution involves an infinite series, and $\sigma$ depends on the aspect ratio $H^{\prime}/W^{\prime}$ ~\cite{jiao2024intrinsic}. In the present case, the transmitting channel corresponds to a rectangular microfluidic conduit described within a 2.5D depth-averaged framework. As a result, no closed-form expression for $\sigma$ is directly applicable. Instead, we evaluate $\sigma$ upon calibration against the simulated permeability of the reference straight channel $\Omega_a$ as
\begin{equation}
    \sigma=\frac{W^2}{k_{\mathrm{simu}}^{\Omega_a}}.
\end{equation}
Doing so ensures exact recovery of the reference state, such that differences among configurations $\Omega_b$--$\Omega_g$ arise solely from the junction-scale correction captured by $H_{\mathrm{eff}}$.  \\

\noindent
Results stemming from our theoretical model are compared with their counterparts based on numerical simulations in Fig.~\ref{Figure5}~(c), with relative deviations always being below $0.3\%$. For instance, we obtain $k_{\mathrm{th}}=37.4$ darcy (to be compared against $k_{\mathrm{simu}}=37.3$ darcy) for configuration $\Omega_b$, while identical values are found for $\Omega_e$ and $\Omega_g$. This level of agreement demonstrates that our effective-aperture formulation accurately captures the junction-scale mechanism controlling channel-scale permeability. \\

\noindent
To provide additional insight, we recast Eq.~(\ref{eq:kth_resistance}) in normalized form upon introducing
\begin{equation}
    \eta=\frac{n_2}{L/W}.
\end{equation}
This represents the fraction of each percolating channel occupied by DEP-TP junctions, thereby quantifying the longitudinal density of DEP junctions along the channel. Accordingly, $n_1 = (1-\eta)L/W$ and $n_2=\eta L/W$. Substituting these expressions into Eq.~(\ref{eq:kth_resistance}) yields
\begin{equation}
    \frac{1}{k_{\mathrm{th}}} = (1-\eta)\frac{1}{k_1} + \eta\frac{1}{k_2}.
\end{equation}
Making use of $k_1 = W_1^2/\sigma$ and $k_2 = H_{\mathrm{eff}}^2/\sigma$, and recalling that the baseline permeability of the reference straight channel is $k_{\Omega_a}= W^2/\sigma$, we obtain
\begin{equation}
    \frac{k_{\mathrm{th}}}{k_{\Omega_a}} = \frac{1}{1-\eta+\eta(W/ H_{\mathrm{eff}})^2}.
    \label{model}
\end{equation}
This provides quantitative support to the conceptual picture according to which permeability enhancement is governed by both the fraction $\eta$ of DEP-TP segments and the effective geometric ratio $W/H_{\mathrm{eff}}$. Note that removing all DEPs, their longitudinal density $\eta$ tends to zero and the system permeability $k_{\mathrm{th}}$  approaches $k_{\Omega_a}$, thus recovering the TPs permeability.

\section{Conclusion}

\noindent
We demonstrate that Dead-End Pores (DEPs), although not directly hosting any flow, exert a significative and systematic influence on the porous system intrinsic permeability. By analyzing a hierarchy of porous structures (ranging from heterogeneous pore spaces to single-pore settings), we isolate the respective roles of DEP geometry and spatial organization (longitudinal density). Our results show that permeability cannot be inferred from DEP volume alone. In fact, the mechanism controlling the local permeability is the fluid-fluid interaction at DEP-TP junctions, which modify the hydraulic resistance. In contrast, individual DEP characteristics (such as depth and orientation) have a negligible impact on the macroscopic flow resistance. \\

\noindent
 We isolate key structural properties significantly controlling the system intrinsic permeability, i.e., ($i$) the effective-aperture of the DEP-TP junction, $H_\mathrm{eff}$ and ($ii$) the longitudinal DEP density, $\eta$. The former is responsible for local enhancement of pore-scale permeability, the latter quantifies its macroscopic spatial extent.
 We then formulate a modeling strategy to effectively capture the impact of DEP on intrinsic permeability. Our theoretical framework is firmly anchored on DEP characteristics and provides a direct, physically grounded connection between pore-scale structure and macroscopic transport properties without the need of additional fitting parameters.\\

\noindent
Overall, our results refine the conventional view of dead-end pores as hydraulically inactive features. Instead, they reveal that DEPs actively modulate flow through localized interactions with the transmitting network. This insight supports the need for a refinement of classical porosity-based models and provides a new perspective for the assessment of permeability in complex porous systems, with implications for a wide range of natural and engineered materials. \\

\section*{ACKNOWLEDGMENTS}
\noindent 
P.d.A. acknowledges the support of FET-Open project NARCISO (ID: 828890) and the Swiss National Science Foundation (grants ID~200021$\_$172587 and 200021$\_$219863). W.J. acknowledges the support of the Agassiz Foundation (ID: 26080102) and Swiss National Science Foundation
(ID: 230535). A.G. acknowledges support from the European Union Next-Generation EU (National Recovery and Resilience Plan - NRRP, Mission 4, Component 2, Investment 1.3 - D.D. 1243 2/8/2022, PE0000005) in the context of the RETURN Extended Partnership. All authors thank Cyprien Soulaine and Radoslav Hurtis for helpful and insightful discussions on simulation methods. \\

\newpage
%\bibliographystyle{unsrt}{}
%\bibliography{library}

\end{document}